\documentclass[aps, prd, twocolumn,a4paper]{revtex4-1}
\usepackage{ulem}
\usepackage{color}
\usepackage{float}
\usepackage{appendix}
\usepackage{graphicx}
\usepackage{epsfig}
\usepackage{epstopdf}
\usepackage{amsmath}
\usepackage{hyperref}
\usepackage{subfig}
\usepackage[dvipsnames]{xcolor}
\newcommand{\be}{\begin{equation}}
	\newcommand{\ee}{\end{equation}}
\newcommand{\ba}{\begin{eqnarray}}
	\newcommand{\ea}{\end{eqnarray}}
\newcommand{\nn}{\nonumber\\}
\begin{document}
\title{Optical properties of an anisotropic hot QCD medium}
\author{M. Yousuf Jamal$^{a,b}$}
\email{mohammad.yousuf@niser.ac.in}
\author{Sukanya Mitra$^{a,c}$}
\email{sukanya.mitra10@gmail.com}
\author{Vinod Chandra$^{a}$}
\email{vchandra@iitgn.ac.in}
\affiliation{Indian Institute of Technology Gandhinagar,  Gandhinagar-382355, Gujarat, India$^a$}
\affiliation{School of Physical Sciences, National Institute of Science Education and Research, HBNI, Jatni-752050, India $^b$}
\affiliation{National Superconducting Cyclotron Laboratory, Michigan State University, East Lansing, Michigan 48824, USA$^c$}

\begin{abstract}
The present investigation involves explorations on the chromo-dielectric properties of the hot QCD medium produced in relativistic heavy-ion collisions in terms of refractive index.
The isotropic/equilibrium modelling is done within an effective quasi-particle model of hot QCD medium. The possibilities of negative refraction in the medium are also explored in terms  of the 
Depine-Lakhtakia index. The anisotropic aspects of the hot QCD medium are incorporated by introducing the anisotropy in a particular direction.
That makes the medium quite similar to uniaxial crystals, and hence we observe phenomenon of birefringence ( two distinct refractive indices in the anisotropic case).
Interestingly, both anisotropy and medium effects play significant roles in deciding the optical properties of the hot QCD/Quark-Gluon-Plasma (QGP) medium.
\end{abstract}
\maketitle

\section{Introduction}
    The hot and dense nuclear matter, Quark-Gluon-Plasma, in relativistic heavy-ion collisions (HIC) behaves more like a near perfect fluid rather than a  non-interacting ultra-relativistic gas of quarks (antiquarks) and gluons~\cite{expt_rhic, expt_lhc, Heinz:2004qz}. The liquidity of the QGP is quantified in terms of the collective flow coefficients such as elliptic flow, $v_2$ and others. The nature of these coefficients in HIC as a function of transverse momentum requires the QGP to be a near perfect fluid with a very small value for the shear viscosity to entropy ($\eta/S$)  (smallest among almost all the known fluids in nature). Apart from the collective behaviour, quarkonia suppression and strong jet quenching ~\cite{Chu:1988wh, Koike:1991mf} are other interesting phenomena associated with the QGP that highlight its plasma aspects (reminiscence of color screening and energy loss).

The goal of the present investigations is to understand plasma aspects of the QGP medium while considering it as a dielectric one and capturing the related interesting physics aspects such as collective plasma excitations and the optical properties such as refractive index of the medium. To understand any medium (say dielectric), the medium must be exposed to the external fields, {\it i.e.,}  electric and magnetic fields. Depending upon its response to the fields, we call it isotropic, anisotropic, linear or non-linear. Once the response of the medium is known in terms of  the permittivity and the permeability, the propagation of the electromagnetic/
chromo-electromagnetic waves could be explored in terms of refractive/ chromo-refractive indices of the medium. 

In the present manuscript, the chromo-refractive index for the hot QCD/ QGP (isotropic as well as anisotropic) medium has been investigated in terms of the chromo-electric permittivity and the chromo-magnetic permeability within semi-classical transport theory.  While setting up the linear transport theory, one must have an adequate modelling of the isotropic/ equilibrium (global/ local) state of the medium. 
The isotropic/ equilibrium state could be described in terms  of the interacting QGP equation of state (EoS) (either computed from the lattice QCD or the improved Hard thermal loop (HTL) resummed perturbation theory).  These EoSs are described  in  terms of  non-interacting system of quasi-gluons and quasi-quarks/antiquarks. 
The starting point is the computation of gluon polarization tensor in the hot QCD/ QGP
medium either within transport theory with the above-mentioned distribution functions or within finite temperature field theory and then extract the responses. As the momentum anisotropy is present in all the stages of the heavy-ion collisions, its inclusion is highly desired while studying any aspects of the QGP.  Keeping this very crucial aspect in mind, we study the response of the anisotropic QGP in the presence of classical chromo-electromagnetic fields, thereby determine optical properties such as dielectric constant and refractive index. In our approach,  the anisotropy could be included at the level of distribution functions by extending the isotropic distribution functions, obtained from the quasi-particle model ~\cite{chandra_quasi1, chandra_quasi2}, in one of the directions (for simplicity one can introduce the anisotropy in the direction of beam propagation). The collective plasma modes within this approach have already been studied  extensively with ideal EoS/ leading order HTL in Refs.~\cite{Romatschke:2003ms, Romatschke:2004jh, Schenke:2006xu, Schenke:2006yp, Carrington:2014bla} and for the interacting EoS within a quasi-particle model in our previous work~\cite{Jamal:2017dqs, Kumar:2017bja}, for both isotropic and anisotropic QGP. The hot QCD/ QGP medium effects were seen to induce significant modifications to the collective modes. 

The prime goal here is to define a chromo-refractive index for the hot QCD medium and search for possibilities of negative refraction (NR) and opacity region (vanishing refractive index) therein for both isotropic and anisotropic cases.
There are a few studies in which the refractive index of weakly as well as strongly coupled plasma has been investigated. Earlier, it was proposed by Veselago~\cite{Veselago:1968}, that the refractive index may be negative in some of the materials. Later on, it was shown ~\cite{Amariti:2010jw} that there is a certain probability of the QGP to have a negative refractive index (NRI) for some frequency range.  Afterwards, there have been various attempts to study the refractive index using a holographic model for strongly coupled plasma ~\cite{Amariti:2011dj, Ge:2010yc, Gao:2010ie, Bigazzi:2011it, Amariti:2011dm}. Juan Liu {\it et al.}~\cite{Liu:2011if}, have studied the refractive index for weakly coupled plasma within HTL perturbation theory. There are several other branches of physics where the researchers have studied the negative refraction and published a lot of informative articles ~\cite{Bond, Jerphagnon, Shao, Lekner, GHOSH, Domanski}.  

In the QCD/ QGP, the basic mathematical quantity, which is needed to understand the response functions, is the gluon polarization tensor. Considering the fact that in the Abelian limit QCD possess the similar features as QED~\cite{Weibel:1959zz}, Jiang {\it et al.} ~\cite{Jiang:2013nw} studied for the first time the refractive index of gluon (chromo-refractive index) in the case of the viscous QGP. They further extended their study of the chromo-refractive index using kinetic theory with a Bhatnagar-Gross-Krook (BGK) collisional kernel ~\cite{Jiang:2016dkf}. It is important to note that, in all the above-mentioned approaches regarding the QGP, the interactions among quarks and gluons in the QGP medium have not been included. Their analysis assumes the QGP as an Ideal gas of gluons, quark and antiquarks (non-interacting, ultra-relativistic). Here, we have incorporated two important aspects while investigating the chromo-refractive index of the QGP medium, {\it viz.}, the EoS effects (via quasi-particle description) and the momentum anisotropy. In the context of the refractive index of the QGP, the former is not yet been included, and the latter is not extensively studied.  As the response of the medium to electric/chromo-electric
and magnetic/chromo-magnetic part is not similar, based on this asymmetry in the electric and magnetic sector, the plasma can be classified as magnetizable and non-magnetizable. Here, the main interest is to explore the magnetizable case and search for the possibility of the refractive index to be negative. As we are working in the Abelian limit (that could be justifiable at higher temperatures where weak coupling results make sense),  for the sake of convenience, we will omit the "chromo" word from now onwards to avoid the repetition.

The paper is organized as follows. In section ~\ref{QOM}, the basic formalism for finding the optical responses and the refractive index has been presented along with the modelling of the isotropic as well as the anisotropic hot QCD medium. In the case of anisotropy, the system shows a quite similar behaviour as uniaxial crystals. We tried to map up some of our observations in the anisotropic medium with the uniaxial crystals. Here, we also discussed the basic formalism of studying negative refraction. Section ~\ref{RandD} deals with results and discussions. The conclusions and the possible future extensions of the present work are offered in section ~\ref{SC}.


\section{QGP as an Optical Medium}
\label{QOM}
The propagation of electromagnetic waves in the QGP medium, treating it as a dielectric one, can be studied in terms of the responses, {\it i.e.,} the permittivity and the permeability of the medium (these can be used  to define the  refractive index of the medium). Before, we compute the refractive index of the  medium, let us first recall, in brief,  the refractive index and its  general aspects. For a wave propagating in a continuous dielectric medium, its electric field vector (${\bf E}$) connects the displacement vector (${\bf D}$) through the electric permittivity tensor($\epsilon_{ij}$) and the magnetic field vector (${\bf B}$) can be expressed in terms of the magnetic field induction (${\bf H}$) through magnetic permeability tensor ($\mu_{ij}$) as,
	\ba
	D_{i}  =  \epsilon_{ij}(\omega,k)E_{j},\  \
	B_{i}  =  \mu_{ij}(\omega,k)H_{j},\label{eq:definition_epsilon_mu}
	\ea
	here $i,j=1,2,3$ are spatial indices, $\omega$ is the frequency and $k \equiv |\mathbf{k}|$ is the propagation vector. In the context of QGP, a covariant treatment of these quantities is required. Therefore, one uses the fluid four-velocity $u^{\alpha}$ to define the four- fields ${\tilde E^{\mu}}$ and ${\tilde B^{\mu}}$ in the Fourier space as,
	\begin{equation}
	\tilde{E}^{\mu}=u_{\alpha}F^{\mu\alpha},\;\tilde{B}^{\mu}=\frac{1}{2}\epsilon^{\mu\rho\alpha\beta}u_{\rho}F_{\alpha\beta},
	\end{equation}
	where $ \epsilon^{\mu\rho\alpha\beta} $ is the four-dimensional Levi-Civita symbols and $\mu$, $\rho$, $\alpha$ and $\beta$ are here Lorentz four-indices (not to confuse with structure functions and magnetic permeability). Using the above equation, one can write the field tensor $F^{\mu\nu}$ as,
	\begin{equation}
	F^{\mu\nu}=\tilde{E}^{\mu}u^{\nu}-\tilde{E}^{\nu}u^{\mu}-\epsilon^{\mu\nu\alpha\beta}\tilde{B}_{\alpha}u_{\beta}.
	\end{equation}
	While including the medium effects, the effective action can be expressed as, 
	\ba
	S_{eff} & = & S_{0}-\frac{1}{2}\int\frac{d^{4}K}{(2\pi)^{4}}A^{\mu}(K)\Pi_{\mu\nu}(K)A^{\nu}(-K)+....\nn
	\label{effea}
	\ea
	Where $A^{\mu}(K)$ is the soft gauge field in momentum space and $K\equiv K_{\mu} = (\omega, k)$. The medium effect is characterized by the polarization tensor $\Pi_{\mu\nu}(K)$. Where the free action($S_{0}$), {\it i.e.,} the action at zero temperature, $T=0$ reads,
	\begin{equation}
	S_{0}=-\frac{1}{2}\int\frac{d^{4}K}{(2\pi)^{4}}\left[\epsilon_{0}\tilde{E}^{\mu}(K)\tilde{E}_{\mu}(-K)-\frac{\tilde{B}^{\mu}(K)\tilde{B}_{\mu}(-K)}{\mu_{0}}\right].
	\end{equation}
	Here, $\epsilon_{0}$ and $\mu_{0}$ are the permittivity and permeability of the free space respectively. In terms of $\epsilon_{\mu\nu}$ and $\mu$, the effective action can be written as,
	\ba
	S_{eff} & = & -\frac{1}{2}\int d^{4}K\Big[\epsilon_{\mu\nu}\tilde{E}^{\mu}(K)\tilde{E}^{\nu}(-K)-\frac{1}{\mu}\tilde{B}_{\mu}(K)\tilde{B}^{\mu}(-K)\Big].\nn
	\label{eq:full_action_weldon}
	\ea 
	One can extract the permittivity and the permeability from the effective action $S_{eff}$ in terms of polarization tensor $\Pi^{\mu\nu}$, which include all the medium effects (comparing  Eq. (\ref{eq:full_action_weldon}) to  Eq.(\ref{effea})). Let us now proceed to the computation of refractive index for the QGP using the response functions($\epsilon(\omega, k)$, $\mu(\omega, k)$).

\subsection{The refractive index for the hot QCD medium}
As stated earlier, the gluon polarization tensor, $\Pi^{\mu\nu}$ is required to define the responses. The formalism of $\Pi^{\mu\nu}$ for interacting EoSs within the quasi-particle description has been presented in details in our previous work ~\cite{Jamal:2017dqs},
and a brief account of that is presented below. To obtain the $\Pi^{\mu\nu}$ in the QCD plasma, we start with an arbitrary particle distribution function, denoting with $f_{i}({\bf p}, X)$ where the index $i$, refers to the particle species (quark, antiquark and gluon). 

Before one proceeds, a particular energy scale is needed to choose in order to study the collective behaviour (plasma behaviour) of hot QCD medium.
We preferred to work on the scale where the collective motion in the hot QCD medium first appears. At this scale, the soft momentum 
$p \sim g T \ll T$, the magnitude of the field fluctuations is of the order of $A \sim \sqrt{g} T$ and derivatives are $\partial_x \sim g T$. This is the scale where one can think of Abelianizing the hot QCD formulations. Note that there have been several attempts to understand the collective behaviour of hot QCD medium either within the semi-classical theory or HTL effective theory ~\cite{Weldon:1982aq,Pisarski:1988vd,Mrowczynski:1993qm, Mrowczynski:1994xv, Mrowczynski:1996vh, Mrowczynski:2000ed, Mrowczynski:2004kv, Mrowczynski:2005ki, Arnold:2003rq}.

In the Abelian limit (omitting the color indices), the space time evolution of the distribution function in the medium  is understood from the Boltzmann-Vlasov ~\cite{Elze:1989un} transport equation below.
\ba
u^{\mu}\partial_{\mu} \delta f^{i}(p,X) + g \theta_{i}
u_{\mu}F^{\mu\nu}(X)\partial_{\nu}^{(p)}f^{i}(\mathbf{p})=\mathcal{C}^{i}(p,X),\nn
\label{eq:Vlasov}
\ea
where , $x^{\mu}=(t,\mathbf{x}) = X$ is the four space-time coordinate and $u^{\mu}=(1,\mathbf{u})$, are the velocity of the plasma particle, respectively with $\mathbf{u}=\mathbf{p}/|\mathbf{p}|$.  $\theta_{i}\in\{\theta_g,\theta_q,\theta_{\bar{q}}\}$ having values $\theta_{g}=\theta_{q}=1$ and $\theta_{\bar{q}}=-1$.   
$\partial_{\mu}$, $\partial_{\nu}^{(p)}$ are the partial
four derivatives corresponding to space and momentum, respectively. $\mathcal{C}^{i}_a(p,X)$, is the collision
term that describes the effects of collisions between
hard particles in a hot QCD medium. We are focusing on the  very near equilibrium case which allows us to neglect the effects from collisions and so  $\mathcal{C}^{i}(p,X) = 0$. The second rank tensor, 
$F_{\mu\nu}$ is the Abelianised chromo-electromagnetic strength tensor which either represents an external field applied to the system, or/and is generated self-consistently by the four-currents present in the  plasma, as follows,
\ba \partial_{\nu}F^{\mu\nu}(x) = J^{\mu}(X),\ea 
where

\ba
J^{\mu}(X) =  g \int  d\varGamma p^{\mu}f({\bf p},X), ~~~~
d\varGamma \equiv \frac{d^{3}p}{(2\pi)^{3}E}. \ea
Eq.(\ref{eq:Vlasov}), can be solved in the linear response approximation {\it i.e.,} the equation can be linearized around the stationary and homogeneous state described by the distribution $f_{i}^{0}({\bf p})$ which is assumed to be neutral when there is no current. The distribution function is then decomposed as:
\ba
f_{i}({\bf p},X) &=& f^{0}_{i}({\bf p}) + \delta f_{i}({\bf p}, X),~~~  f^{0}_{i}({\bf p}) \gg \delta f_{i}({\bf p}, X).\nn    
\ea

Next,  the current, $J^{\mu}$, induced by a soft gauge field, $A^{\mu}$ can be obtained in terms of the $\delta f_{i}({\bf p},X)$ as,
\ba
J^{\mu}_{\rm ind}(X) =  g \int  d\varGamma p^{\mu} \delta f({\bf p}, X).
\ea
Where $\delta f({\bf p},X)$,  contains the fluctuating part and given as
\ba
\delta f({\bf p},X) = 2 N_{c} \delta f_{g}({\bf p},X) + N_{f} (\delta f_{q}({\bf p},X) - \delta f_{\bar{q}}({\bf p},X)).\nn
\ea
Here, $\delta f_{g}({\bf p},X)$, $\delta f_{q}({\bf p},X)$ and $\delta f_{\bar{q}}({\bf p},X)$ 
are the fluctuating parts of the gluon, the quark and antiquark densities respectively.
After solving the transport equations for the fluctuations 
$\delta f_{g}$ and  $\delta f_{q/\bar{q}}$,  we get the induced current in the Fourier space as,
\ba
J'^{\mu}_{\rm ind}(K) &=& g^2 \int \frac{d^3 p}{(2\pi)^3} u^{\mu}\partial^{\beta}_{(p)} f({\bf p}) \bigg[ g_{\alpha \beta}- \frac{u_{\alpha} K_{\beta}}{K\cdot u + i \epsilon}\bigg] A'^{\alpha}(K),\nn
\label{eq:induced current}
\ea
where, $\epsilon$ is a very small parameter needed to avoid unwanted infinities and will be sent to zero in the end.
The distribution function, $f({\bf p})$ in terms of isotropic/equilibrium quark/antiquark (at zero baryon density, $f_q\equiv f_{\bar{q}}$) and gluon distribution functions is given as,
	\begin{equation}
	\label{fq}
	f({\bf p}) = 2 N_c f_{g}({\bf p})+ N_{f} (f_{q}({\bf p})+f_{\bar{q}}({\bf p})).
	\end{equation}
	In the linear approximation, the equation of motion for the gauge field can be obtained in Fourier space as,
	\begin{equation}
	J'^{\mu}_{ind}(K) = \Pi^{\mu\nu}(K)A'_{\nu}(K). 
	\label{eq:linear induced current}
	\end{equation}
	The gluon selfenergy tensor is symmetric and transverse in nature, {\it i.e.,}  $\Pi^{\mu\nu}(K) = \Pi^{\nu\mu}(K)$ and follows the Ward's identity, $K_{\mu}\Pi^{\mu\nu}(K) = 0$. 
	From Eq.(\ref{eq:induced current}) and Eq.(\ref{eq:linear induced current}) we can obtain $\Pi^{\mu\nu}(K)$ as,  
	\begin{equation}
	\Pi^{\mu\nu}(K) = g^{2}\int \frac{d^{3}p}{(2\pi)^{3}} u^{\mu}\frac{\partial f(p)}{\partial p^{\beta}}\bigg[g^{\nu\beta} - \frac{K^{\beta} u^{\nu}}{K\cdot u + i\epsilon}\bigg].
	\label{iso_pi}
	\end{equation} 
Next,  we shall discuss in brief the effective fugacity quasi particle model (EQPM) that has been employed here to incorporate the medium interaction effects.

\subsection{The  EQPM}    
	As we mentioned earlier, the isotropic modelling of the medium is done within a quasi-particle description of hot QCD medium. To that end, we employed EQPM where the medium effects have been encoded in the effective gluon and effective quark fugacities ~\cite{chandra_quasi1, chandra_quasi2}. These fugacity parameters define effective gluon and quark/antiquark momentum distribution functions for the isotropic/equilibrated medium. There are several other quasi-particle models where the medium modifications are captured in terms of effective mass for gluons and quarks~\cite{effmass1, effmass2, polya}. Others are NJL and PNJL based effective models~\cite{pnjl} and recent quasi-particle models based on the Gribov-Zwanziger (GZ) quantization results leading to non-trivial IR-improved dispersion relation in terms of Gribov parameter ~\cite{flor}.
		 These quasi-particle models have shown their utility while studying the transport properties of the 
	QGP~\cite{Bluhm,chandra_eta, chandra_etazeta, PJI, Mkap, Mitra:2017sjo}. 
		In  Ref.~\cite{Greco}, the ratio of electrical conductivity to shear viscosity has been explored within the framework of the effective mass model. 
		As mentioned above, the present analysis considers the EQPM for the investigations on the optical properties of the hot and dense medium produced in HIC experiments.
		Mitra and  Chandra~\cite{Mitra:2016zdw} computed the electrical conductivity and charge diffusion coefficients within EQPM.
		 In the context of quarkonia physics~\cite{Chandra:2010xg, Agotiya:2016bqr}  and thermal particle  production ~\cite{Chandra:2015rdz, Chandra:2016dwy}  and heavy quark transport~\cite{Das:2012ck, Chandra:2015gma} too, the EQPM played an important role.
		 There are issues with the above mentioned other approaches while comparing the transport coefficients with their phenomenological estimates~ \cite{Ryu, Denicol1} from experimental observables at RHIC. 
	Nevertheless, these quasi-particle approaches serve the purpose of modelling the equilibrated/ isotropic state of the QGP which is crucial for the transport theory computations. 
		The hot QCD equations of state (EoSs) described here in terms of EQPM are
	the very recent (2+1)-lattice EoS from hot QCD collaboration ~\cite{bazabov2014} (LEoS), and the 3-loop HTL perturbative EoS that has recently been computed by N. Haque {\it et,  al.} ~\cite{nhaque, Andersen:2015eoa} which agrees reasonably well with the recent lattice results~\cite{bazabov2014,fodor2014}. 
	
	\begin{figure}
		\centering
		\includegraphics[height=6.30cm,width=9.60cm]{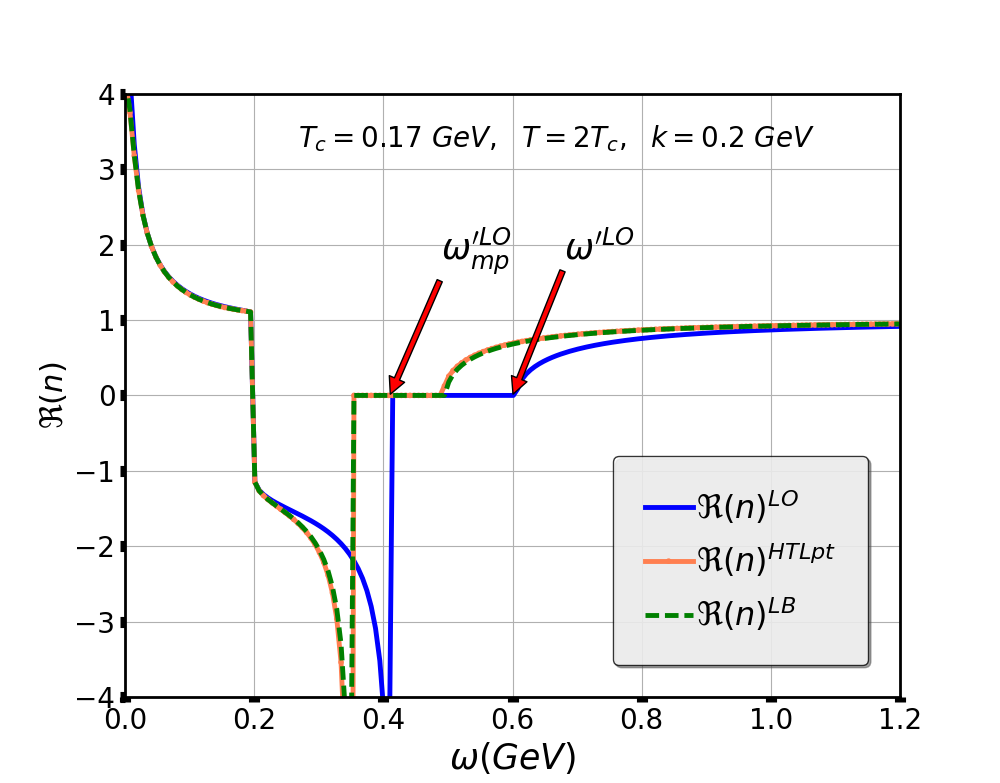}
		\caption{$\Re(n)$ is plotted for isotropic case ($\xi = 0$) considering various EoSs.}
		\label{fig:rn_iso}
	\end{figure}
	\begin{figure}
		\centering
		\includegraphics[height=6.30cm,width=9.60cm]{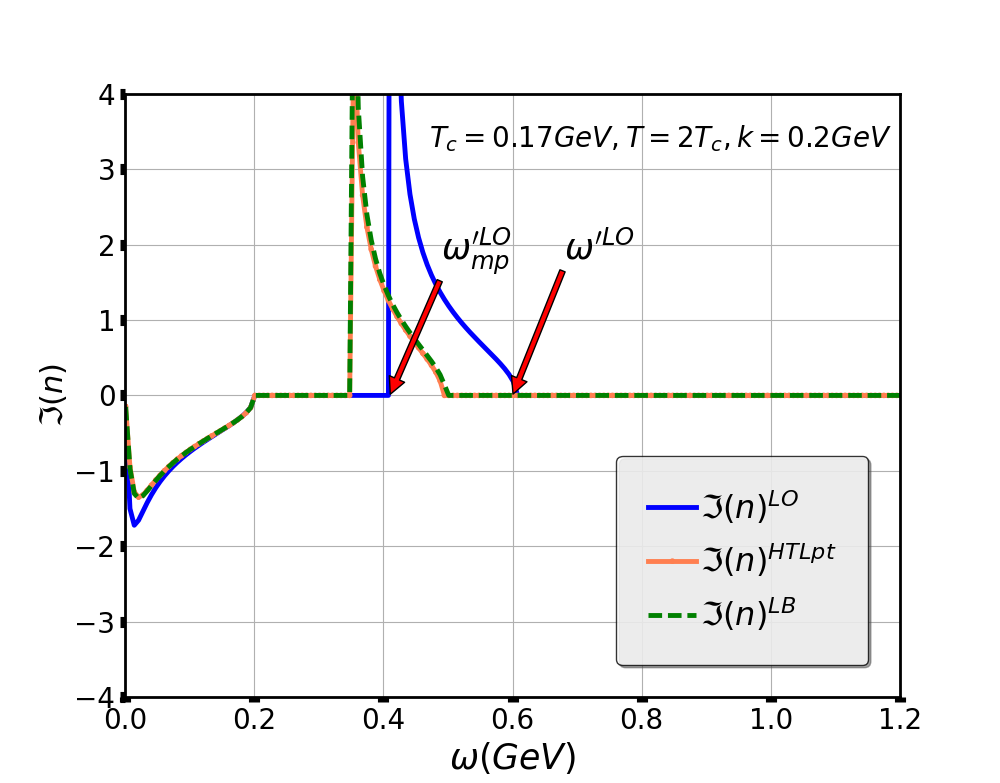}
		\caption{$\Im(n)$ is plotted for isotropic case ($\xi = 0$) considering various EoSs.}
		\label{fig:in_iso}
	\end{figure}
	
	\begin{figure}
		\centering
		\includegraphics[height=6.30cm,width=9.60cm]{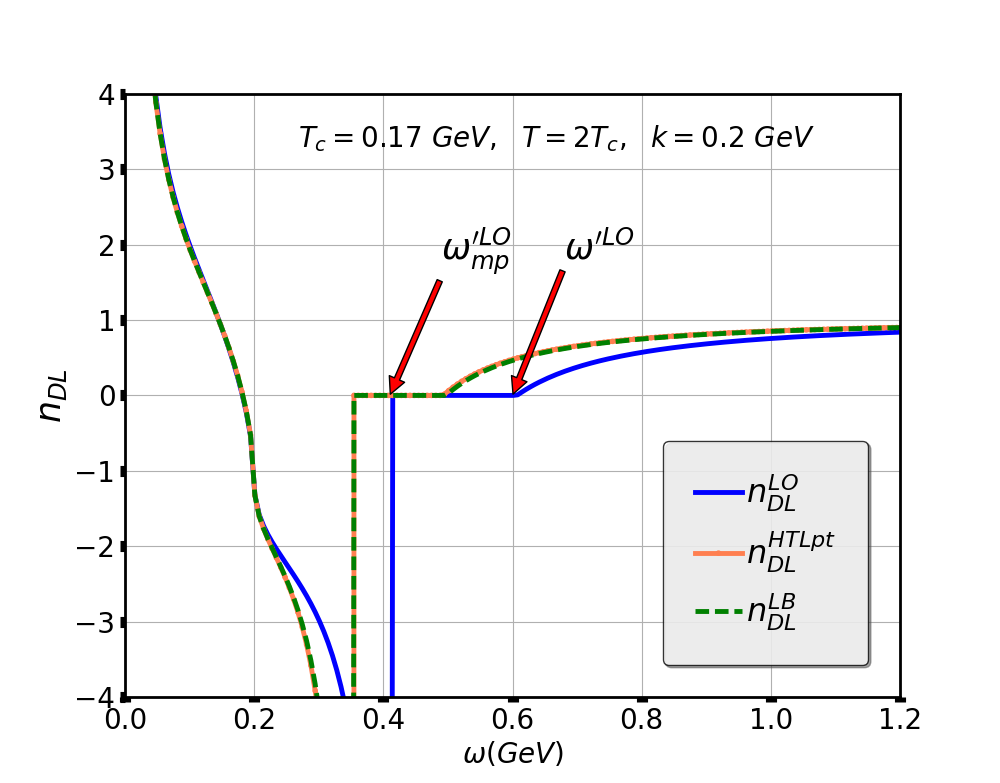}
		\caption{$n_{DL}$ is plotted for isotropic case ($\xi = 0$) considering various EoSs.}
		\label{fig:ndl}
	\end{figure}
	
	The EQPM serves as the input in terms of the effective  quasi-particle distribution functions, $ f_{eq}\equiv \lbrace f_{g}, f_{q} \rbrace$  (describing the strong interaction effects in terms of effective fugacities $z_{g,q}$) as ~\cite{chandra_quasi1, chandra_quasi2},
	\ba
	\label{eq1}
	f_{g/q}= \frac{z_{g/q}\exp[-\beta E_p]}{\bigg(1\mp z_{g/q}\exp[-\beta E_p]\bigg)},
	\ea
	where, $\beta = 1/T$, $T$ is the temperature in energy units. $E_{p}=|{\bf p}|$ for the gluons and $\sqrt{|{\bf p}|^2+m_q^2}$ for the quark degrees of freedom ($m_q$ denotes the mass of the quarks). This leads to the following dispersion relation, 
	\ba
	\label{eq2}
	\omega_{g/q}=E_{p}+T^2\partial_T ln(z_{g/q}).
	\label{epp}
	\ea
	
	The effective fugacities, $z_{g,q}$ are obtained from the above mentioned EoS by realizing the the hot QCD medium as Grand canonical system of quasi-gluons and quasi-quarks/antiquarks with momentum distributions given in Eq. ~(\ref{eq1}) as follows. We denote the effective partition function for the hot QCD medium by $Z=(Z_g~\times~Z_q)$, $Z_g$ for gluon and $Z_q$ for quarks. The corresponding expressions in terms of $z_g$ and $z_q$ are as follows,
	\ba
	\ln (Z_g) &=& -\nu_g~V \int \frac{d^3p}{8\pi^3}\ln(1-z_g \exp(-\beta p)), \nn
	\ln (Z_q) &=& \nu_q~V \int \frac{d^3p}{8\pi^3}\ln(1+z_q \exp(-\beta p)),\nn
	\ln(Z)&=&\ln (Z_g)+\ln (Z_q),\label{eq:par}
	\ea
	where the gluonic degrees of freedom, $\nu_g=2(N_c^2-1)$ and the quark degrees of freedom, $\nu_q=2~\times~2~\times~N_c~\times~N_f$ for $SU(N_c)$.
	Now using the thermodynamic relation, $P\beta V=\ln (Z)$, we can match the right-hand side of Eq.~ \ref{eq:par}, with the lattice data for the pressure for ($2+1$)-flavor QCD. Here, $P$ denotes the pressure and $V$ denotes the volume. 
	From this relation, we can numerically obtain the temperature dependence of $z_g$ and $z_q$ (for more details please see Ref.\cite{chandra_quasi2}).
	Considering the pure gluonic case, the temperature dependence of fugacity parametrs,  $z_g$ has been determined using the relation,
		\ba
	P_g = -\frac{\beta^{-4}\nu_g}{2\pi^2}\int_{0}^{\infty}w^2\ln(1-z_g\exp(-w))dw,
		\ea
	and the temperature dependence of $z_q$ is obtained using the following relation,	
		\ba
	(P-	P_g) = \frac{\beta^{-4}\nu_g}{2\pi^2}\int_{0}^{\infty}w^2\ln(1-z_q\exp(-w))dw,
		\ea
		where $w(w=\beta p)$ is a dimensionless quantity.
	
	One further requires, Debye mass ($m_D$) computed within EQPM and effective QCD coupling (depicting charge renormalization) in the medium. To compute these quantities, we follow our previous works ~\cite{Mitra:2016zdw, Jamal:2017dqs} and references therein. Based on that one can obtain the $m_D^2$ as,
	\ba
	\label{dm}
	m_D^2&=& -4 \pi \alpha_{s}(T) \bigg(2 N_c \int \frac{d^3 p}{(2 \pi)^3} \partial_p f_g ({\bf p})\nn&+&2 N_f \int \frac{d^3 p}{(2 \pi)^3} \partial_p f_q ({\bf p})\bigg),
	\ea
	where, $\alpha_{s}(T)$ is the QCD running coupling constant at finite temperature ~\cite{qcd_coupling} and described in detail in the context of EQPM in~\cite{Mitra:2016zdw}.
	\begin{figure*}
		\centering
		\includegraphics[height=5.80cm,width=18.80cm]{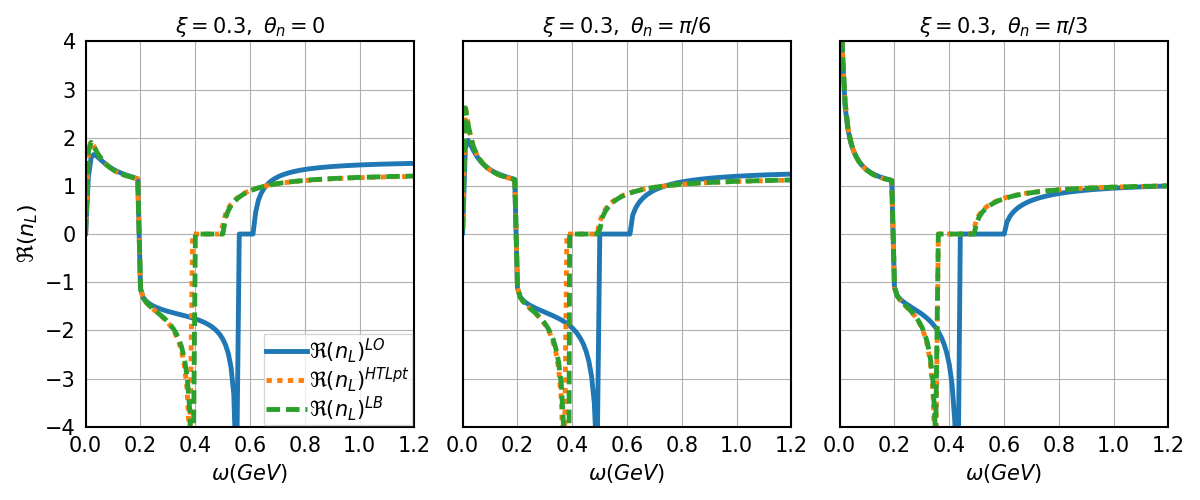}
		\includegraphics[height=5.80cm,width=18.80cm]{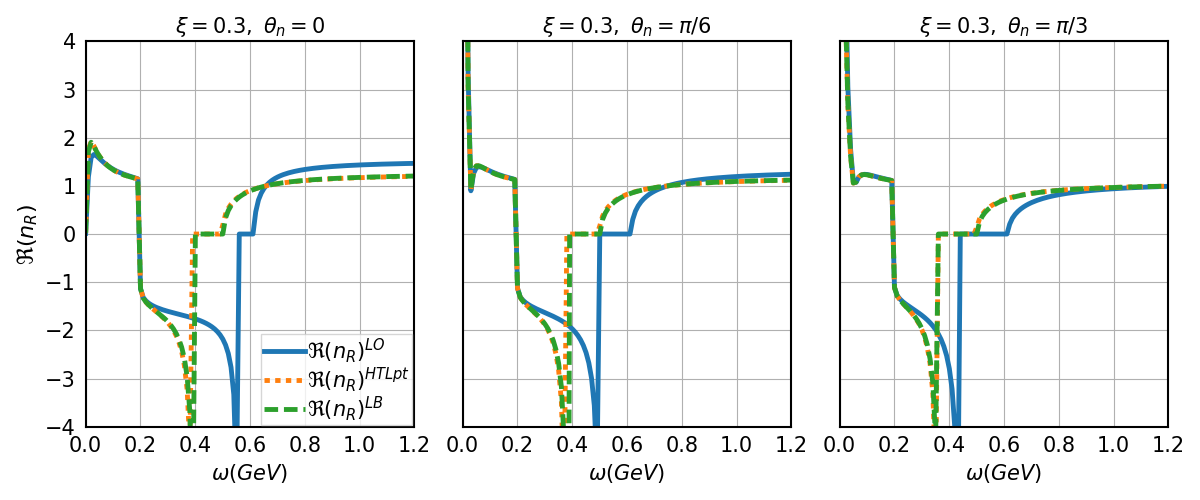}
		\caption{Real part of $n_L$ and $n_R$ for various EoSs at $ k =0.2~GeV$, $T_{c} = 0.17~GeV$ and $T=2T_c$ at different angles.}
		\label{fig:re_nl_nr}
	\end{figure*} 

\subsection{Response Functions for anisotropic hot QCD medium}
	To describe the anisotropic hot QCD medium, we follow the approach employed in Ref. ~\cite{Jamal:2017dqs, Kumar:2017bja, Romatschke:2003ms, Kobes:1990dc, Romatschke:2004jh}. In this approach,  the anisotropic momentum distribution functions for the gluons and quark/ antiquarks are obtained by rescaling (stretching and squeezing) of one direction in momentum space distribution function as,
	\ba
	f_{\xi}({\bf\tilde{p}}) = C_{\xi}~f(\sqrt{{\bf p}^{2} + \xi({\bf p}\cdot{\bf \hat{n}})^{2}}).
	\ea
	This introduces  one more degree of freedom,  {\it viz. }, the direction of anisotropy, ${\bf \hat{n}}$ with ${\bf \hat{n}}^{2} = 1$.
	The anisotropy parameter   $\xi$  can be adjusted to reflect either squeezing ($\xi > 0$) or stretching $(-1<\xi<0$)  of the distribution in the ${\bf \hat{n}}$ direction. As mentioned earlier, this makes the system resembles with the uniaxial crystals where there is one preferred direction (one optic axis). Choosing, $\xi = 0$ will bring us to the case of the isotropic QCD medium which shows quite similar results as the isotropic crystals in the context of optical properties. The normalization, $C_{\xi}$ is fixed by demanding the expression for the 
	$m_D(T)$ to be the same in both isotropic and anisotropic medium. This leads to
	\ba
	C_{\xi} = 1+\frac{\xi}{3} +\mathcal{O}(\xi^{2}).
	\ea
	Next, performing the change of variables  
	$({\bf |\tilde p|}\equiv \sqrt{{\bf p}^{2} + \xi({\bf p}.{\bf \hat{n}})^{2}} )$ in Eq.(\ref{iso_pi}) and considering the temporatl axial gauge, we get
	\ba
	\Pi^{ij}(K)&=&m_D^2~ C_{\xi}\int\frac{d\Omega}{4\pi}u^i\frac{u^l+\xi(\mathbf{u}\cdot\mathbf{\hat{n}})n^l}{(1+\xi(\mathbf{u}\cdot\mathbf{\hat{n}})^2)^2} \bigg[\delta^{jl}\nn 
	&+&\frac{u^jk^l}{\omega+-\mathbf{k}\cdot\mathbf{u}}\bigg],
	\label{aniso_pi}
	\ea
	where, $m_D$ from Eq.~\ref{dm}is obtained by employing EQPM as,
	\ \ba
	\label{dm1}
	{m_D^2}^{(EoS)}&=&4 \pi \alpha_{s}(T) T^2  \bigg( \frac{2 N_c}{\pi^2} PolyLog[2,z_g^{(EoS)}]\nn&-&\frac{2 N_f}{\pi^2} PolyLog[2,-z_q^{(EoS)}]\bigg), 
	\ea
	and contains effects from various EoSs where $z_{g/q}\rightarrow1$, corresponds to LO or ideal EoS. 
	\begin{figure*}
		\centering
		\includegraphics[height=5.80cm,width=18.80cm]{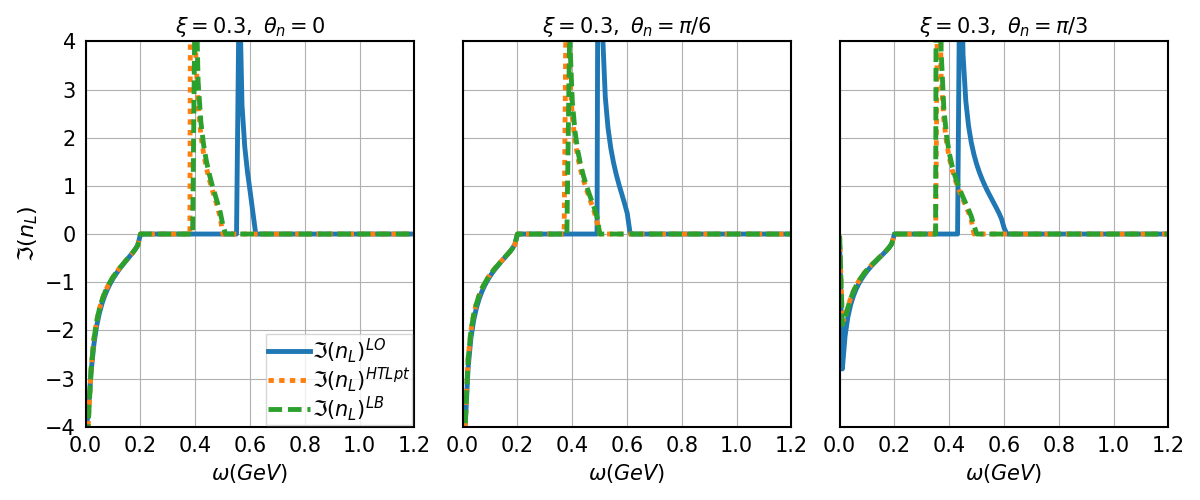}
		\includegraphics[height=5.80cm,width=18.80cm]{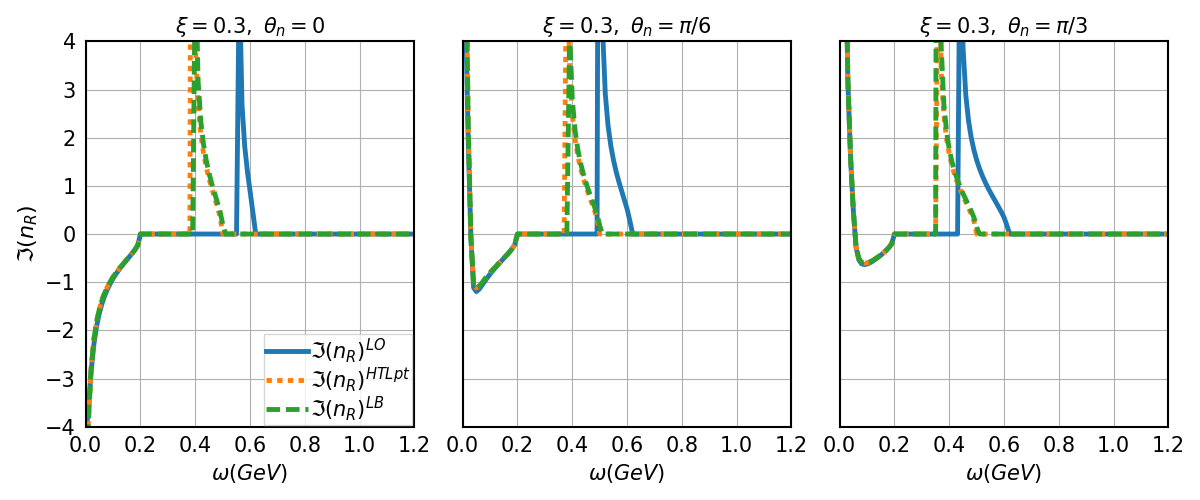}
		\caption{Imaginary part of $n_L$ and $n_R$ for various EoSs at $ k =0.2~GeV$, $T_{c} = 0.17~GeV$ and $T=2T_c$ at different angles.}
		\label{fig:im_nl_nr}
	\end{figure*}
	
	To solve (\ref{aniso_pi}) a tensor decomposition of $\Pi^{ij}(\omega, k)$ one need to construct an analytic form using available symmetric tensor and then perform a suitable contraction. For isotropic case, we need only two, the transverse,
	$P^{ij}_{T}=\delta^{ij}-k^{i}k^{j}/{k^{2}}$ and the longitudinal,
	$P^{ij}_{L}=k^{i}k^{j}/{k^{2}}$ tensor projectors to decompose, $\Pi^{ij}$. While considering the anisotropy into account with
	anisotropy vector ${\bf \hat{n}}$, we have to consider two more
	projectors $P^{ij}_{n}={\tilde{n}}^i{\tilde{n}}^j/{\tilde{n}}^2$ 
	and $P^{ij}_{kn}=k^{i}{\tilde{n}}^{j}+k^{j}{\tilde{n}}^{i}$. Where, $\tilde{n}^i=(\delta^{ij}-\frac{k^i k^j}{k^2})\hat{n}^j$
	is a vector orthogonal to, $k^i$ ,{\it i.e.,}   $\mathbf{\tilde{n}}\cdot{\mathbf{k}}=0$. Thus, we can decompose the gluon selfenergy given in Eq.~(\ref{seexpan}) into following four basis as follows,
	\ba
	\Pi^{ij}=\alpha{P}^{ij}_{T}+\beta{P}^{ij}_{L}+\gamma{P}^{ij}_{n}+\delta{P}^{ij}_{kn},
	\label{seexpan}
	\ea
	where 
	$\alpha$, $\beta$, $\gamma$ and $\delta$, are called as structure functions and can be determined by taking the appropriate contractions of $\Pi^{ij}$ as,  
	\ba
	\alpha&=&({P}^{ij}_{T}-{P}^{ij}_{n})\Pi^{ij}, ~~~\beta={P}^{ij}_{L}\Pi^{ij} \nonumber,\\
	\gamma&=&(2{P}^{ij}_{n}-{P}^{ij}_{T})\Pi^{ij},~~~\delta=\frac{1}{2 k^{2}{\tilde{n}}^2}{P}^{ij}_{kn}\Pi^{ij}.
	\label{structurefunctions}
	\ea 
	Applying these contractions in Eqs. ~(\ref{aniso_pi}) and ~(\ref{seexpan}) and comparing them, one can obtained the structure functions in the limit of small anisotropy as,
{\small	\ba
	\alpha &=&C_{\xi}\bigg[\frac{m_D^2 z^2}{2}+\frac{m_D^2 \xi }{48 k}\bigg\{ \bigg(-2 k \left(15 z^4-19 z^2+4\right) \cos (2 \theta_n )\nn&-&2 k \left(9 z^4-9 z^2+4\right)\bigg)+ \bigg(\bigg(3 k \left(z^2-1\right) \left(5 z^2-3\right) z \cos (2 \theta_n )\nn
	&-&3 k \left(z^2-1\right) z+9 k \left(z^2-1\right) z^3\bigg)-\frac{1}{4} m_D^2 z \left(z^2-1\right)\bigg)\nn&\times&\log \left(\frac{z+1}{z-1}\right)\bigg\}\bigg],
	\ea}
	
{\small	\ba
	\beta &=&C_{\xi}\bigg[ m_D^2 \left(-z^2\right) \bigg(\xi  \cos ^2(\theta_n )-\frac{2 \xi }{3}-\frac{1}{2} \xi  z^2 (3 \cos (2 \theta_n )+1)\nn
	&+&\frac{1}{4} z \log \left(\frac{z+1}{z-1}\right) \left(\xi  \left(3 z^2-2\right) \cos (2 \theta_n )+\xi  z^2-2\right)+1\bigg)\bigg]\nn
	\ea}
	and
{\small	\ba
	\gamma &=&C_{\xi}\frac{ m_D^2 }{12}\xi \left(z^2-1\right) \sin ^2(\theta_n ) \bigg(6 z^2-4\nn
	&-&3 \left(z^2-1\right) z \log \left(\frac{z+1}{z-1}\right)\bigg),
	\ea}
	with $z=\frac{\omega }{k}$.
	
	Using $\Pi^{\ij}(\omega, k)$ in Eq.(\ref{eq:full_action_weldon}) for the anisotropic hot QCD/ QGP medium, we obtained two eigen values of permittivity,
	\ba
	\epsilon_{L} &=& 1-\frac{\beta'}{K^{2}}
	\label{eq:aniso_eps_L}
	\ea
	and
	\ba
	\epsilon_{R} = 1-\frac{\beta'}{K^{2}} - \frac{\gamma}{\omega^2}.
	\label{eq:aniso_eps_R}
	\ea
	The permeability from the same is obtained as,
	\ba
	\frac{1}{\mu(\omega,k)} & = & 1+\frac{K^{2}\alpha-\omega^{2}\beta'}{k^{2}K^{2}},
	\label{eq:aniso_magnetizable_mu}
	\ea
	where, $\beta'$ is defined as,
	\ba
	\beta'&=&\beta-C_{\xi} \frac{m_D^2}{12 k^3} \bigg(2 k^3 (\xi -6)-6 k \xi  \cos (2 \theta_n ) \left(k^2-3 \omega ^2\right)\nn
	&-&3 \omega  \log \left(\frac{2 k}{\omega -k}+1\right)\Big(\xi  \cos (2 \theta_n ) \left(3 \omega ^2-2 k^2\right)-2 k^2\nn
	&+&\xi  \omega ^2\Big)+6 k \xi  \omega ^2\bigg).
	\ea
	
	For real valued $\omega$ the structure functions are real for all $\omega > k$ and complex for $\omega < k$.
	For imaginary values of $\omega$, all four structure functions are real.  Since all the structure constants depend on the Debye mass($m_{D}$), any modification in Debye mass will modify all of them.
	
	It is to note that, $\epsilon_{L}$ and $\epsilon_{R}$ overlap in the case when anisotropy is not considered as, $\gamma =0$ when $\xi=0$.  Now, having permittivity and permeability, one can analyze the refractive index $n$ as,
	\ba
	n(\omega, k) = \sqrt{\epsilon(\omega, k) \mu(\omega, k)}.
	\label{eqref}
	\ea

The QGP as a dissipative medium, is expected to have complex refractive index,  $n(\omega, k) = \Re(n) +i\ \Im(n)$.
In the other branches of physics also, it has been observed that a medium can have a complex refractive index ~\cite{Mead, Adair, Glauber}. 
The real part leads to  the phase velocity, (${\bf v_p} = \frac{1}{\Re(n)}{\bf{\hat k}}$). For $\Re(n)<1$, the phase velocity is greater than the speed of light in the medium, and if $\Re(n)<0$,  the direction of propagation will be opposite to the direction of phase velocity. 
On the other hand, the $\Im(n)$ corresponds to the attenuation, whenever an electromagnetic wave traverses through the medium. In general, one has $\Im(n)>0$, which corresponds to the absorption of the wave and $\Im(n)=0$  shows no loss in the medium whereas  $\Im(n)<0$, indicates the amplification of the wave in the medium.  As we shall see later,  all these three cases for the $\Im(n)$ could be realized in the case of the anisotropic QGP. 
Notably, in the case of anisotropic medium, one can define two polarization states of the  medium that could be realized in terms of  $\epsilon_L$ and $\epsilon_R$, leading two different refractive indices denoted as, $n_{L}$  and $n_{R}$, respectively and defined as,
\ba
n_{L/R}(\omega, k) = \sqrt{\epsilon_{L/R}(\omega, k) \mu(\omega, k)}.
\label{eq:nLR}
\ea
Here, $n_L$ has found to have a similar behaviour whereas $n_R$ behaves slightly differently as that of the isotropic case (discussed in details in the results section). 
Below is a brief discussion on how one can relate this scenario from the optics (optical crystals) point of view. The mediums can also be classified as being either isotropic or anisotropic based on their optical behaviour. Isotropic mediums have equivalent axes that interact with the electromagnetic wave in the same manner, regardless of their orientation with respect to the incident wave. 
Whereas, the anisotropic mediums have a non-uniform spatial/momentum distribution, which results in different values being obtained when they are probed from different directions. 
The wave entering the isotropic medium is refracted at a constant angle and passes through the medium with a single velocity without being polarized.  While the wave enters the anisotropic medium having one preferred direction ( optic axis of the crystals and in our case the beam axis or the direction of anisotropy ${\hat n}$), 
it gets polarized and is refracted into two ways, each travels at different velocities and experience different refractive indices, depending upon the direction of propagation. 
This phenomenon is known as birefringence or double refraction. This has been extensively studied from the optics point of view and several excellent published articles are present in the literature ~\cite{Bond, Jerphagnon, Shao, Lekner, GHOSH, Domanski}.   
One among the two waves, passing through the anisotropic medium, obeys the laws of normal refraction, and travels with the same velocity in every direction through the medium, {\it i.e.,} it behaves similarly as that of the isotropic case. 
Generally, in optical physics, this wave is called the ordinary wave/ray. The other wave travels with a velocity that is dependent upon the propagation direction within the medium termed as the extraordinary wave/ray. 
\begin{figure*}
	\centering
	\includegraphics[height=5.50cm,width=18.80cm]{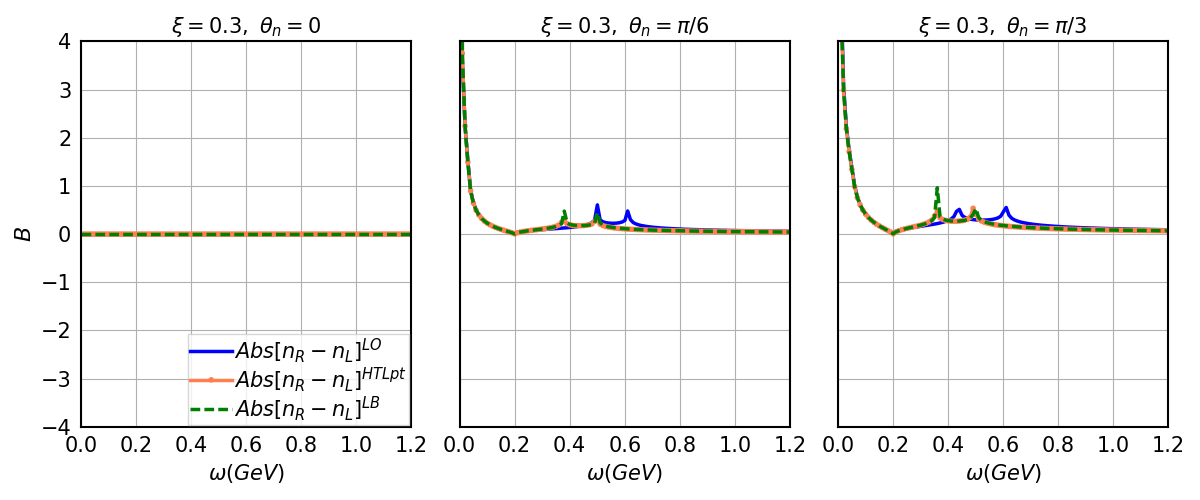}
	\caption{Birefringence for various EoSs at $ k =0.2GeV$ $T_{c} = 0.17GeV$ and $T=2T_c$ at different angles.}
	\label{fig:BN}
\end{figure*}
If the incident chromo-electromagnetic wave impact the medium in a direction which is parallel to the preferred direction, both behave as an ordinary wave and do not separate into two. In the present case, this is same if one takes $\theta_n=0$ {\it i.e., } the propagation of the particle in the direction of anisotropy.
The optical path lengths of the waves emerging from the medium are identical, and hence, there is no relative phase shift. One can observe this by calculating the birefringence (B) of the waves travelling through the anisotropic medium as,
\ba
B = |n_o-n_e| = |n_L-n_R|,
\label{eq:br}
\ea
which is expected to be zero when considering in or parallel to the preferred direction of the medium (or the optic axis).
Here,  $n_o$ and $n_e$ are ordinary and extraordinary refractive indices respectively. 
As the values for each component of refractive index can vary, the absolute value in Eq.~\ref{eq:br} can determine the total amount of birefringence. 
Fixing the other parameters one can get birefringence number for a wave of a particular frequency, $\omega$. 
Let us now proceed to discuss the negative refraction that has been studied in several branches of physics with various observables~\cite{Foteinopoulou, astel1, astel, Sikes}.

\subsection{Description of Negative Refractive Index(NRI)}
\label{sec:NRI}
The quadratic nature of definition of refractive index, {\it i.e.,} $n^2 = \epsilon(\omega, k)\ \mu(\omega, k)$, implies that
it is not sensitive to the sign of $\epsilon(\omega, k)$ and $\mu(\omega, k)$. It was proposed by Veselago \cite{Veselago:1968} that the
simultaneous change of sign of permittivity ($\epsilon(\omega,k)$) and permeability ($\mu(\omega,k)$) corresponds to a crossover between
different branches of the square root, {\it i.e.,} from $n=\sqrt{\epsilon(\omega, k) ~ \mu(\omega, k)}$ to $n = -\sqrt{\epsilon(\omega, k)
~ \mu(\omega, k)}$, or from the positive refractive index to the negative one. Next, if the medium is dissipative, the $\epsilon$ and $\mu$,
and  the $n$, are complex quantities. The condition for the NRI both in isotropic and anisotropic QGP is as follows. Whenever $\Im(\mu)$ and
$\Im(\epsilon)$ vanish and both $\ Re(\mu)$ and  $\ Re(\epsilon)$ are negative simultaneously, in a particular frequency range, in this case,
NRI can be realized in the medium. In the following discussion, we shall see that for the isotropic medium the sign of $\epsilon(\omega, k)$
and $\mu(\omega, k)$ have a significant physical implication.
	
Note that the phase velocity is defined by
\begin{equation}
\mathbf{v}_{p}=\frac{1}{\mathrm{\Re}(n)}\hat{\mathbf{k}}=v_{p}\hat{\mathbf{k}},
\end{equation}
whose sign is the same as that of $\mathrm{Re}(n)$. But the direction of the energy flow or the Poynting vector is not affected by the sign 
of $\epsilon$ and $\mu$. In a medium with small dissipation, the direction of the energy flow coincides with that of the group velocity,
\begin{equation}
	\mathbf{S}=v_{g}U\hat{\mathbf{k}},
	\end{equation}
where, $U$ is a positive time-averaged energy density and $v_{g}=d\omega/dk$. 
If we have a negative phase velocity, the direction of the phase velocity can be opposite to the energy flow or the group velocity, ({\it i.e.,}
the direction of phase velocity can be radially inward and the direction of energy flow can be radially outward or vice-versa). 
\begin{equation}
	v_{p}<0,\; v_{g}>0.
	\end{equation}
	Employing this  a better criterion for the NRI in the isotropic medium has been derived, called the Depine-Lakhtakia index ($n_{DL}$)~\cite{DL}, 
	\begin{equation}
	n_{DL}=\left|\epsilon\right|\mathrm{\Re}(\mu)+\left|\mu\right|\mathrm{\Re}(\epsilon).
	\label{eq:nDL}
	\end{equation}
	Whenever $n_{DL}<0$, we have NRI region, and also the directions of the phase velocity and the energy flow are opposite.  This is a better indicator for the NRI in the medium that one can use to investigate the isotropic medium and has been widely used in various branches of physics~\cite{Akhlesh, Lakhtakia:2004zk, Tom, Akyurtlu, Cheng:2018zjv}.  This condition works well for the isotropic expansion of the medium as there is a spherical symmetry but perhaps does not holds good for the anisotropic case because of the asymmetry in the medium.  For the case of anisotropy in the medium, this method can be employed at extremes, {\it i.e.,} for the parallel, $\theta_n = 0$ and perpendicular, $\theta_n=\pi/2$ cases. Therefore, in the anisotropic case, we shall employ the former criterion once it is checked for the isotropic case with $n_{DL}$.

\section{Results and Discussions}
\label{RandD}
The expressions obtained in the previous sections have been plotted and shown in different figures to observe their variation with respect to the frequency ($\omega$) and their angular dependence. In this section, we shall discuss the important features of these plots. All the figures have been accommodated with the results coming from all three EoSs, {\it viz}, recent (2+1)-lattice, 3-loop HTL perturbative and non-interacting ideal EoSs and denoted, respectively as LB, HTLpt and LO at the temperature $T=2T_c$.

In the isotropic case, the real and the imaginary part of the refractive index, $(n)$ as well as the Depine-Lakhtakia index, $n_{DL}$ have been obtained considering, $\xi=0$ and shown, respectively,  in Fig.~\ref{fig:rn_iso},\ref{fig:in_iso} and \ref{fig:ndl}. It has been found that in the frequency range $\omega < k$, both the $\Re(n)$ and the $n_{DL}$ are positive and so the refractive index is positive whereas, in this region, $\Im(n)$ is negative which corresponds to the amplification of the wave. For the frequencies, $k \leq \omega \leq \omega^{'}_{mp}$ (where $\omega^{'}_{mp}$ is the frequency at which $\mu$, given in Eq.\ref{eq:aniso_magnetizable_mu}, is having a pole), the $n_{DL}$ and the $\Re(n)$ are negative while $\Im(n)$ is zero and hence, this is the region of negative refraction while there is no dissipation in this range of frequency. In the range from $\omega^{'}_{mp}$ to $\omega^{'}$ (to avoid the bulk, $\omega^{'}_{mp}$ is only shown for LO case in the plots), the $n_{DL}$ and the $\Re(n)$ vanish while $\Im(n) > 0$, and hence, the medium is highly dissipative and opaque for the chromo-electromagnetic waves. For $\omega = \omega^{'}$ and onwards, $\Im(n) = 0$, whereas $n_{DL}$ and the $\Re(n)$ are positive and approaching to unity. That shows there is a normal refractive index, but the phase velocity is greater than the speed of light.

It is important to note that the frequency ranges where the refractive index is negative, {\it i.e.,} $ \omega \in (k,\omega^{'}_{mp}) $ and  where the medium is opaque, {\it i.e.,} $\omega \in (\omega^{'}_{mp} , \omega^{'})$ for the chromo-electromagnetic wave are proportional to $m_D(T)$, which has different values for different EoSs at constant temperature. Therefore, the interaction effects that entered through the  Debye mass modified the results which are observed in the ideal case. In fact, the medium interaction effects, through $m_D(T)$, have been observed to reduce the frequency range of opacity and the negative refraction.

Next, to understand the same properties in the presence of anisotropy in the hot QCD/QGP medium, we considered the anisotropy in a particular direction, $ {\bf \hat{n}}$. The results are plotted for the real and imaginary parts of $n_{L/R}$ with the anisotropic strength, $ \xi = 0.3$ using Eq.~(\ref{eq:nLR}), in  Fig.~\ref{fig:re_nl_nr} and Fig.~\ref{fig:im_nl_nr} respectively. 
Here,   we observed two distinct refractive indices, which is quite similar to the case of birefringence in uniaxial crystals. Note that, in the uniaxial crystals, there is one preferred direction, {\it i.e.,} the optical axis which in our case can be thought of as the beam axis or the direction of anisotropy ${\hat n}$. If one considers the propagation of a particle in the direction of anisotropy {\it i.e.,} $\theta_n =0$, both the refractive indices turns out to be same. 
This can be seen in the left panel of Fig.~\ref{fig:BN}, where we have plotted birefringence for various angular dependence using Eq.~\ref{eq:br}. While in the middle and right panel of the same figure, we observe a finite degree of birefringence that indicates the presence of the two distinct refractive indices.

For different angles ($\theta_n=0,\pi/6~ \text{and}~ \pi/3$) as shown in first row of Fig.~\ref{fig:re_nl_nr} and Fig. ~\ref{fig:im_nl_nr}, it has been observed that the $\Re(n_L)$ and the $\Im(n_L)$ follow a similar pattern as we found in the isotropic case while the $\Re(n_R)$ and the $\Im(n_R)$ while 
with  a slightly different pattern at the smaller frequencies. In this context, if one relates this phenomenon with the birefringence of crystals, one can say that $n_L$ behaves as the ordinary refractive index while $n_R$ as the extraordinary refractive index ( depends on the direction of propagation of the particle in the medium).
For smaller range of frequency both $\Re(n_R)$ and $\Im(n_R)$ are positive. As the frequency is  slightly higher, the $\Im(n_R)$ becomes negative while $\Re(n_R)$ remains positive and at $\omega = k$, both vanish.  Afterwards, they follow a similar pattern as for the isotropic case with some quantitative differences in the numbers.
Finally, the medium effects incorporated by employing the EQPM (using both Lattice and HTLpt) showed a similar behaviour as it was in the ideal case (LO) but modified the observed ranges, in each case considered here.

\section{Summary and Conclusions}
\label{SC}
In conclusion, the chromo-electromagnetic response functions of the anisotropic hot QGP/QCD, in the Abelian limit, have been investigated and their implication in understanding the dielectric properties of the hot QCD/QGP medium have been studied. In the current approach, the hot QCD medium interactions are incorporated exploiting the quasi-particle description of the hot QCD EoS either computed within recent lattice QCD methods or 3-loop HTL perturbation theory.

Realizing the hot QGP as an optical medium with the refractive index, we observe a frequency region where the refractive index could be negative and also an opacity region where the real part of the refractive index vanishes. In this context, we analyzed the Depine-Lakhtakia index for the isotropic medium and studied its behaviour in detail to probe the directions of energy flow there. On the other hand, for the anisotropic case depending on the direction of the anisotropy {\it i.e.,} for $\theta_n \neq 0$, the presence of momentum anisotropy in the medium causes the split of the eigenstate of the dielectric tensor. Hence, we obtained two different eigenvalues as $\epsilon_L$ and $\epsilon_R$ and two independent refractive indices. Although, in the case when $\theta_n = 0$, these two eigenvalues overlap even in the presence of anisotropy. These conditions mimic the response of the uniaxial crystal to the electromagnetic waves. Therefore, we discuss in details the birefringence which a widely adopted phenomenon studied in optics. The possibility of negative refraction in the anisotropic case turned out to be quite tricky. Defining the Depine-Lakhtakia index is perhaps not a
very viable way to understand this very crucial phenomenon.  
Nevertheless, we looked at the regions where the conditions for the negative refraction based on the nature of  $\epsilon$ and $\mu$, discussed in section ~\ref{sec:NRI}, are satisfied. In both the cases (isotropic or anisotropic medium), the refractive indices, get significant contributions from the hot QCD medium effects as compared to the case when QGP is approximated as the ideal system (non-interacting ultra-relativistic gas of quarks, antiquarks and gluons).  Furthermore, the results after incorporating the hot QCD medium effects through different EoSs followed the similar pattern but shifted the frequency ranges and modified observed regions of the opacity and the negative refraction. 
These effects basically narrowed these ranges as compared to the non-interacting ideal case. 

An important question that may arise regarding the significance of the present work to the experimental observables for the QGP at HIC. To that end, we can utilize the analysis to compute the polarization energy loss in the hot QCD/QGP medium and finally relate it with the jet quenching 
(nuclear modification factor, $R_{AA}$). Apart from that, the NRI and the opacity aspects could perhaps be related to the photon production phenomenon in HIC. 
This is the matter of immediate future extension of the work.  Another interesting future extension
of the work would be to look at the contributions from the collisions and collectivity and near perfect fluidity of the hot QCD medium/QGP and study its impact on collective plasma properties and refractive index. 
Equally importantly, obtaining an expression for the inter-quark potential in this medium and its phenomenological aspects in terms of understanding the quarkonia dissociation in the hot QCD medium will be another exciting direction where our future investigations will focus on.

\section*{Acknowledgements}
 V. Chandra would like to sincerely acknowledge DST, Govt. of India for Inspire Faculty Award -IFA13/PH-15 and Early Career Research Award (ECRA) Grant 2016. M. Y. Jamal would like to thank NISER Bhubaneswar for providing postdoctoral position. We would like to acknowledge people of INDIA for their generous support for the research in fundamental sciences in the country.
  
{}
\end{document}